 \documentclass[aps,amsmath,amssymb]{revtex4}
     \usepackage{graphicx}
     \usepackage{epsfig}

\begin{document}

\title{Vlasov analysis of relaxation and meta-equilibrium}
\author{Celia Anteneodo}
\address{Departamento de F\'{\i}sica, Pontif\'{\i}cia 
Universidade Cat\'olica do Rio de Janeiro, \\
CP 38071, 22452-970, Rio de Janeiro, Brazil, and \\
Centro Brasileiro de Pesquisas F\'{\i}sicas, \\
         R. Dr. Xavier Sigaud 150,  
         22290-180, Rio de Janeiro, Brazil \\
Email: celia@cbpf.br } 
\author{Ra\'ul O. Vallejos}
\address{Centro Brasileiro de Pesquisas F\'{\i}sicas,\\
         R. Dr. Xavier Sigaud 150,  
         22290-180, Rio de Janeiro, Brazil \\
Email: vallejos@cbpf.br}

%%%%%%%%%%%%%%%%%%%%%%%%%%%%%%%%%%%%%%%%%%%%%%%%%%%%%%%%%%%%%%%%%%%%%

\begin{abstract}
%\abstracts{
The Hamiltonian Mean-Field model (HMF), an inertial $XY$ ferromagnet 
with infinite-range interactions, has been extensively studied in 
the last few years, especially due to its long-lived meta-equilibrium 
states, which exhibit a series of anomalies, such as, breakdown of ergodicity, 
anomalous diffusion, aging, and non-Maxwell velocity distributions.
The most widely investigated meta-equilibrium states of the HMF arise 
from special (fully magnetized) initial conditions that evolve to a 
spatially homogeneous state with well defined macroscopic 
characteristics and whose lifetime increases with the system size, 
eventually reaching equilibrium. 
These meta-equilibrium states have been observed for specific energies 
close below the critical value 0.75, 
corresponding to a ferromagnetic 
phase transition, and disappear below a certain energy close to 0.68.
In the thermodynamic limit, the $\mu$-space dynamics is governed by a Vlasov equation. 
For finite systems this is an approximation to the exact dynamics. However, it 
provides an explanation, for instance, for the violent initial relaxation and 
for the disappearance of the homogeneous states at energies below 0.68.
%}
%
\end{abstract}

\maketitle

%%%%%%%%%%%%%%%%%%%%%%%%%%%%%%%%%%%%%%%%%%%%%
     \section{Introduction}
Consider the one-dimensional Hamiltonian
\begin{equation} \label{ham0}
H   =  \frac{1}{2} \sum_{i=1  }^N     p_{i}^{2} +
       \frac{J}{2N} \sum_{i,j=1}^N 
       \left[   1-\cos(\theta_{i}-\theta_{j}) \right] \; .
\end{equation}
It represents a lattice of classical spins with infinite-range interactions.
Each spin rotates in a plane and is therefore described by an angle 
$-\pi \le \theta_i < \pi$, and its conjugate angular 
momentum $p_i$, with $i=1,\ldots,N$;  the constant $J$ is the interaction strength. 
Of course, one can also think of point particles of unitary mass moving on a circle.
This model is known in the literature as mean-field XY-Hamiltonian (HMF)\cite{antoni95}.  

The HMF has been extensively studied in the 
last few years (see \cite{reviewHMF} for a review). 
The reasons for such interest are various.    
From a general point of view, the HMF can be considered the simplest 
prototype for complex, long-range systems like galaxies and plasmas 
(in fact, the HMF is a descendant of the mass-sheet gravitational 
model\cite{antoni95}). But the HMF is also interesting for its anomalies, 
be them model-specific or not.
Especially worth of mention are the long-lived meta-equilibrium 
states (MESs) observed in the ferromagnetic HMF. 
These states exhibit breakdown of ergodicity, anomalous diffusion, and non-Maxwell 
velocity distributions, among other anomalies\cite{metastable,plr} (see also the contribution  
by A. Rapisarda et al. in this volume). 
It has been conjectured that it may be possible to
give a thermodynamic description of these MESs by extending the standard statistical 
mechanics along the lines proposed by Tsallis\cite{tsallis}.

The simplicity of the HMF makes possible a full analysis of its equilibrium 
statistical properties, either in the canonical\cite{antoni95} or microcanonical 
ensembles\cite{antoni02}.
If interactions are attractive ($J>0$), the system exhibits a ferromagnetic 
transition at the critical energy $E_c=0.75JN $. 
Here we will focus on the out-of-equilibrium behavior of the ferromagnetic HMF 
($J>0$), when the system is prepared 
in a fully magnetized configuration, at an energy close below $E_c$, with 
uniformly distributed momenta (``water-bag'' initial conditions). 
Under these initial conditions the system evolves to a 
spatially homogeneous state with well defined macroscopic 
characteristics and whose lifetime increases with the system size, 
eventually reaching equilibrium. 
Numerical experiments have shown the 
disappearance of the family of homogeneous MESs 
below a certain energy close to $0.68JN$.

%%%%%%%%%%%%%%%%%%%%%%%%%%%%%%%%%%%%%%%%%%%%%
     \section{Equations of motion}
It is convenient to write the Hamiltonian (\ref{ham0}) in the simplified form:
$$  %%\label{hamsimp}
H   =    \frac{1}{2} \sum_{i=  1  }^N p_{i}^{2} + \frac{N}{2} \left( 1-m^2 \right),
$$
where we have introduced the magnetization per particle 
$$
{\bf m} =   \frac{1}{N} \sum_{i=  1}^N \hat{\bf r}_i  \; ,
\;\;\;\;\; \mbox{with} \;\;\;
\hat{\bf r}_i =   (\sin \theta_i, \cos \theta_i) \; ,
$$
and for simplicity we have taken $J=1$. The equations of motion read
$$  %%\label{motioneqs}
\dot{\theta}_i \;=\;\frac{\partial H}{\partial p_i}\,=\,p_i, 
\;\;\;\;\;\;\; 
\dot{p}_i \;=\;-\frac{\partial H}{\partial \theta_i} \,=\,{\bf m} \cdot \hat{\bf \theta}_i,
$$
for $i=1,\ldots,N$, with $ \hat{\bf \theta}_i=( \cos \theta_i,-\sin \theta_i)$.
Without loss of generality, we can set the axes such that $m_x(t=0)=0$. 
If, additionally,  the distribution of momenta is symmetrical, then $m_x(t)=0,\;\;\forall t$. 
In that case, the equations of motion become 
$$ %%\label{singlex}
\dot{\theta}_i \;=\;\frac{\partial H}{\partial p_i}\,=\,p_i, 
\;\;\;\;\;\;\; 
\dot{p}_i \;=\;-\frac{\partial H}{\partial \theta_i} \,=\, - m(t) \sin \theta_i.
$$
Notice that these equations can be seen 
as the equations for a pendulum with a time-dependent length. 

%%%%%%%%%%%%%%%%%%%%%%%%%%%%%%%%%%%%%%%%%%%%%
    \section{First stage of relaxation}
Fully magnetized states violently relax to a state of vanishing magnetization, 
within finite size corrections. 
The most elementary approach to describing the relaxation of ${\bf m}$, from a given initial 
condition, is to perform a series expansion around $t=0$, i.e.,  
$$
 {\bf m}(t) = \sum_{k\geq 0} \frac{1}{k!} {c_k} \,t^k .
$$
In our case,  the initial condition is such that $m=1$ (with $m_x=0$), 
then one obtains the following coefficients for $m_y(t)$ 
\begin{eqnarray} \nonumber
c_0 &=& 1 \\  \nonumber
c_2 &=& -\langle p^2 \rangle_0  \\ \nonumber 
c_4 &=& \langle p^4 \rangle_0  +4\langle p^2 \rangle_0 \\ \nonumber
c_6 &=& -\Bigl( \langle p^6 \rangle_0 +26\langle p^4 \rangle_0 +16\langle p^2 \rangle_0 
+18\langle p^2 \rangle^2_0 \Bigr)\\ \nonumber
& \vdots & 
\end{eqnarray}
and $c_{\rm odd}=0$, where averages are calculated with the initial distribution of momenta $h(p)$. 
If $h(p)$ at $t=0$ is symmetrical around $p=0$, then $m(t)$ is an even function of $t$. 
In particular, if the initial condition is water-bag, i.e., 
$\theta_i=0, \forall i$ and additionally $p_i$ are uniformly distributed in the 
interval $[-p_o,p_o]$, then (from Hamiltonian (\ref{ham0}),  
$p_o=\sqrt{6\varepsilon}$, with $\varepsilon$ 
the energy per particle), one obtains 
\begin{equation} \label{exact}
m_y(t)\;=\;1-\varepsilon t^2+\biggl( \frac{3\varepsilon^2}{10} 
+\frac{\varepsilon}{3} \biggr)t^4 +\ldots \;.
\end{equation}
The convergence of this series is very slow and, given that a general expression 
is not available, only the very short 
time of the relaxation can be described.

%%%%%%%%%%%%%%%%%%%%%%%%%%%%%%%%%%%%%%%%%%%%%
       \section{Vlasov equation}
On the other hand, the evolution equation of the reduced probability density function (PDF) in  
$\mu$-space is formally equivalent to the Vlasov-Poisson system\cite{antoni95}
\begin{equation} \label{vlasov0}
\frac{\partial f}{\partial t} \,+\, p\frac{\partial f}{\partial \theta}
\,-\, \frac{\partial V}{\partial \theta}\frac{\partial f}{\partial p}\,=\,0,
\end{equation}
where 
$V=-{\bf m}\cdot \hat{\bf r}(\theta)$ 
and
${\bf m}=\int d\theta \,\hat{\bf r}(\theta)\int dp \,f(\theta,p,t)$.
If ${\bf m}= m \hat{y}$, then
\begin{equation}  \label{vlasovx}
\frac{\partial f}{\partial t} \,+\, p\frac{\partial f}{\partial \theta}
\,-\, m\sin\theta\frac{\partial f}{\partial p}\,=\,0,
\end{equation}
with
\begin{equation} \label{my}
 m(t) =\int_{-\pi}^{\pi} d\theta \,\cos \theta  \int_{-\infty}^\infty dp \,f(\theta,p,t).
\end{equation}
The Vlasov equation (\ref{vlasovx}) can be cast in the form
$$
\frac{\partial f}{\partial t}  \;=\; \left[ { L}_0 +{ L}_1(t) \right] f ,
$$
where ${ L}_0= -p\partial_\theta$ and ${ L}_1(t)= m(t)\sin\theta\partial_p$.
We will consider states (for instance, with vanishing magnetization) 
for which the term $L_1(t)$ can be treated as a perturbation. 
It is convenient to switch to the interaction representation, i.e., 
to define $\widetilde{f}(t)\;=\; {\rm e}^{ -{L}_0 t} f(t)$, then
$$
\frac{\partial \widetilde{f}}{\partial t}  \;=\; \widetilde{ L}_1(t) \widetilde{f} ,
$$
where $\widetilde{L}_1={\rm e}^{ -{ L}_0 t} { L}_1 {\rm e}^{ { L}_0 t}$.

The equation for the propagator $\widetilde{U}$, such that 
$\widetilde{f}(t)=\widetilde{U}(t)\widetilde{f}(0)$, is
$\partial \widetilde{U}/\partial t  = \widetilde{ L}_1\widetilde{U}$, 
therefore, 
$$
\widetilde{U}(t)  \;=\; 1 \;+\; \int_0^t dt'\widetilde{ L}_1(t')\widetilde{U}(t') ,
$$
and recursively, one has
\begin{equation}  \label{propagator}
\widetilde{U}(t)  \;=\;1 
+ \int_0^t dt_1\widetilde{ L}_1(t_1) 
+ \int_0^t dt_1\widetilde{ L}_1(t_1) \int_0^{t_1} dt_2\widetilde{ L}_1(t_2)
+ \ldots \;.
\end{equation}
The solution at order $k$ of the Vlasov Eq. (\ref{vlasovx}) is
\begin{equation} \label{solk}
f^{(k)}(\theta,p,t)\;=\; {\rm e}^{ -pt\partial_\theta} \widetilde{U}^{(k)}(t)   f(\theta,p,0),
\end{equation}
where the index $(k)$ indicates the order at which the expansion (\ref{propagator}) is truncated.
From here on, we will deal with continuous distributions, hence our treatment is valid in the 
thermodynamic limit.

%%%%%%%%%%%%%%%%%%%%%%%%%%%%%%%%%%%%%%%%%
     \subsection{Lowest-order truncation} \label{zero}
At zeroth-order, the propagator is approximated by 
$\widetilde{U}\simeq \widetilde{U}^{(0)}=1$. This is equivalent to 
neglecting the magnetization. Thus, if ${\bf m}=0$, the truncation is exact. 

For the initial distribution $f(\theta,p,0)=  g(\theta) h(p)$, where $g(\theta)$ is uniform in 
$[-\pi,\pi]$ (hence, ${\bf m}=0$) and $h(p)$ is an arbitrary even function, 
both distributions remain unaltered in time, consistently with the numerical simulations in 
Fig. 7 of \cite{plr}. In fact, if ${\bf m}=0$ for any time, 
there are no forces to drive the system out of the macroscopic state.
 
If the initial condition is $f(\theta,p,0)=  \delta(\theta) h(p)$, where $\delta$ is the  
Dirac delta function and $h(p)$ an arbitrary even function of $p$ (as in our 
case of interest), although $\bf m\neq 0$, $L_1$ is small (it is null at $t=0$ 
and remains small for later times), allowing a perturbative treatment. Then, we have
$$
f^{(0)}(\theta,p,t) \;=\; {\rm e}^{-pt\partial_\theta} f(\theta,p,0) 
                    \;=\; h(p) \,\delta(\theta-pt)  
  \;=\; \frac{1}{2\pi} h(p) \, \sum_{k=-\infty}^{\infty} {\rm e}^{ik(\theta-pt)} .
$$
Therefore, 
$$
h^{(0)}(p,t) \;=\; \int_{-\pi}^\pi d\theta f^{(0)}(\theta,p,t) \;=\; h(p),
$$
that is, at zeroth-order, the distribution of momenta, whatever it is, 
does not change in time. 
However the angular distribution does indeed change. 
For instance, 
in the particular case of the water-bag distribution, where 
$h(p)$ is a uniform distribution in $[-p_o,p_o]$, we obtain 
\begin{equation} \label{g0}
g^{(0)}(\theta,t) \;=\; \int_{-\infty}^\infty dp f^{(0)}(\theta,p,t)  \;=\; 
\frac{1}{2\pi}\left( 1+2 \sum_{k\geq 1} \cos(k\theta) \frac{\sin( kp_0 t)}{k p_0 t} \right).
\end{equation}
Notice that  the distribution of angles becomes uniform in the long time limit. 
It gets uniform through a mechanism of phase mixing, where particles do not interact (remember 
that magnetization has been neglected). 
From Eqs. (\ref{my}) and (\ref{g0}), the zeroth-order magnetization is 
\begin{equation} \label{m0order}
m^{(0)}(t) \;=\; \int_{-\pi}^\pi d\theta \cos\theta \, g^{(0)}(\theta,t) 
\;=\;\frac{\sin( p_0 t)}{p_0 t}, 
\end{equation}
whose expansion in powers of time yields
$$
m^{(0)}(t)  \;=\; 1-\varepsilon t^2+\frac{3}{10}\varepsilon^2 t^4 +\ldots
$$
Observe that this  expansion up to second-order coincides with the exact 
one, given by Eq. (\ref{exact}), for any $\varepsilon$.

%%%%%%%%%%%%%%%%%%%%%%%%%%%%%%%%%%%%%%%%%
   \subsection{First-order truncation}
Now,  recalling that 
$\widetilde{U}^{(1)}=1+\int_0^tdt_1\widetilde{ L}_1(t_1)$, from (\ref{solk}), 
at first-order, we have
\begin{eqnarray} \nonumber
f^{(1)}(\theta,p,t) &=& {\rm e}^{-pt\partial_\theta}  
\widetilde{U}^{(1)}(t)  f(\theta,p,0)  \\ \nonumber
                    &=& f^{(0)}(\theta,p,t) +                      
{\rm e}^{-pt\partial_\theta}\int_0^t dt_1 \widetilde{ L}_1(t_1) f(\theta,p,0),
\end{eqnarray}
where  $\widetilde{ L}_1(t)=
{\rm e}^{pt\partial_\theta}\biggl( m(t)\sin\theta\,\partial_p \biggr)
{\rm e}^{-pt\partial_\theta}$. 
Then
\begin{eqnarray} \nonumber
f^{(1)}(\theta,p,t) \;=\; h(p)\delta(\theta-pt) &+&
h'(p)\delta(\theta-pt)\int_0^t dt_1 m(t_1) \sin(\theta+pt_1-pt) \\ \nonumber
&-& h(p)\delta'(\theta-pt)\int_0^t dt_1 \,t_1 m(t_1) \sin(\theta+pt_1-pt) \;.
\end{eqnarray}

\begin{figure}[htb]
\begin{center}
\includegraphics*[bb=100 420 550 670, width=0.7\textwidth]{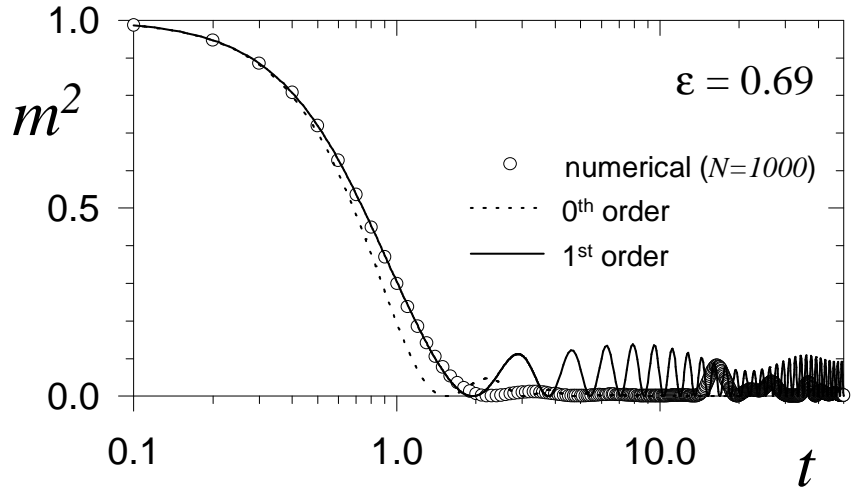}
\caption{\protect 
Squared magnetization as a function of time. 
The initial state is fully magnetized with uniformly distributed momenta for $\varepsilon=0.69$. 
Symbols correspond to numerical simulations with $N=1000$, the distribution of momenta is regular. 
Dashed lines correspond to the zeroth-order approximation obtained from Eq. (\ref{m0order}).
Full lines correspond to the first-order approximation given by Eq. (\ref{m1order})
} 
\end{center}
\label{fig:magnet}
\end{figure}

\noindent
After some algebra, for the case $h(p)$ uniform in $[-p_o,p_o]$, we obtain
\begin{equation}
\label{m1order}
m^{(1)}(t) \;=\; m^{(0)}(t)  \;+\;  \int_0^t dt_1 \, \frac{m(t_1) }{2p_0}
 \biggl(  \sin(p_o[t-t_1]) +   \frac{t_1-t}{t_1+t} \sin(p_o[t+t_1])   \biggr)
\end{equation}
Substituting $m(t)$ by $m^{(0)}(t)$ one obtains the magnetization at first-order.
Moreover,
\begin{eqnarray} \nonumber
h^{(1)}(p,t) \;=\; \int_{-\pi}^\pi d\theta f^{(1)}(\theta,p,t) \;=\; h(p) 
&+& h'(p)\int_0^t dt_1 m(t_1) \sin(pt_1) \\ \nonumber
&+& h(p)\int_0^t dt_1 \,t_1m(t_1) \cos(pt_1) \; .
\end{eqnarray}
We recall that, for the uniform distribution, $h'(p)\propto [\delta(p+p_o)-\delta(p-p_o)]$. 
This explains why $h(p,t)$ presents two spikes at $p=\pm p_o$.

Fig.~1 shows the first stage of the relaxation of the magnetization. 
Numerical simulations were performed for $N=1000$. Increasing the system size does not 
change the numerical curve in the time  interval considered ($t\le50$).  
Of course, for longer times the curve becomes size dependent\cite{metastable,plr}
The squared magnetization rapidly decreases from its initial value $m^2=1$ down to zero at 
$t\simeq 2$. Then, it remains very close to zero up to $t\simeq20$. From then on, one 
observes bursts of small amplitude. 
Since Vlasov equation is exact in the thermodynamic limit, it describes the exact $N=1000$ 
evolution up to time $t\simeq 50$. 
The zero order approximation describes correctly $m^2$ vs $t$ for a very short time 
($t\simeq 0.4$). 
The first-order approximation describes satisfactorily the violent initial relaxation 
(up to $t\simeq 2$), but it does not reproduce the structure appearing later. 
Higher order corrections are required to describe that behavior. 
Extrapolation of numerical simulations\cite{metastable,plr} shows that $m\to0$ in the 
thermodynamic limit. This regime settles for times beyond the scope of our approximation.

%%%%%%%%%%%%%%%%%%%%%%%%%%
   \subsection{Equilibrium}

For completeness,  let us discuss the distributions at thermal equilibrium\cite{vlasov_eq}. 
If the system has already attained equilibrium, then $\partial_t=0$. 
Let also assume that the equilibrium distribution can be factorized, i.e.,
$f(\theta,p)\;=\;g(\theta)\,h(p)$.
Then, from (\ref{vlasovx}),
\begin{equation} \label{id1}
p\frac{\partial g}{\partial \theta} \,h(p) \;=\; 
m\sin\theta\,g(\theta) \frac{\partial h}{\partial p}.
\end{equation}
Assuming  $h(p) = { A} \exp(-\beta p^2/[2\mu])$, 
Eq. (\ref{id1}) reduces to 
$\partial g/\partial \theta = -m\sin\theta\,g(\theta)$. 
Thus
\begin{equation} \label{id3}
g(\theta) \;=\; C {\rm e}^{\beta m\cos\theta} ,
\end{equation}
with the normalization constant $C=1/[2\pi I_0(\beta  m)]$, 
where $I_0$ is the modified Bessel function of zeroth-order. 
The equilibrium magnetization can be obtained from the consistency condition (\ref{my}):
$$
m\;=\;\int_{-\pi}^\pi d\theta  \cos\theta \,g(\theta)
\;=\;\frac{I_1(\beta m)}{I_0(\beta m)},
$$
thus recovering the results of canonical calculations\cite{antoni95}.

%%%%%%%%%%%%%%%%%%%%%%%%%%%%%%%%%%%%%%%%%%%%%%
   \subsection{Meta-equilibrium}

Although we have not found the long-time solution of Vlasov equation, starting 
from fully magnetized initial conditions, numerical simulations\cite{metastable} 
indicate that in the 
thermodynamic limit the system tends to a spatially homogeneous state. 
We have seen in Sect. \ref{zero} that, once reached a homogeneous state, the 
distribution of momenta, whatever it is, does not change in time. 
But, the question is whether the  homogeneous
solutions are stable or not under perturbations. 
On one hand, the Vlasov approach is a good approximation to the discrete 
dynamics, on the other finite-size effects may be the source of perturbations 
that may take the system out of a Vlasov steady state. 
Therefore, we will perform a stability test (valid in the thermodynamic limit) 
and discuss the results under the light of the discrete dynamics.

There is the well known Landau analysis\cite{balescu}
which concerns {\em linear} stability. 
A more powerful stability criterion for homogeneous equilibria has been proposed by 
Yamaguchi et al.\cite{yamaguchi}. This is a nonlinear criterion specific to the  
HMF. It states that $f(p)$ is stable if and only if the quantity 
\begin{equation}
I\,=\,1\,+\,\frac{1}{2} \int_{-\infty}^\infty {\rm d}p\, \frac{f^\prime(p)}{p}
\end{equation}
is positive (it is assumed that $f$ is an even function of $p$).
This condition is equivalent to the zero frequency case 
of Landau's recipe\cite{antoni95,choi03}. 
Yamaguchi et al.\cite{yamaguchi} showed that a distribution which is spatially 
homogeneous and Gaussian in momentum becomes unstable below the transition energy
 $\varepsilon_{\rm cr}=3/4$ (see also \cite{choi03,inagaki93}), 
in agreement with analytical and numerical results for finite $N$ systems. 
They also showed that homogeneous states with zero-mean uniform $f(p)$ 
are stable above $\varepsilon=7/12=0.58...$ (see also \cite{antoni95,choi03}).
In the same spirit, it is instructive to analyze the stability of the family 
of $q$-Gaussian distributions 
\begin{equation} \label{qgaussian}
f(p) \propto \exp_q(-\alpha p^2) =\left[ 1- \alpha (1-q) p^2 \right]^{1/(1-q)} \; ,
\end{equation}
which allows to scan a wide spectrum of PDFs, from finite-support 
to power-law tailed ones, containing as particular cases the Gaussian ($q=1$) and the
water bag ($q=-\infty$). In Eq. (\ref{qgaussian}), the normalization constant has 
been omitted and the parameter $\alpha>0$ is related to the second moment 
$\langle p^2 \rangle$, which is finite only for $q<5/3$. 
In the homogeneous states of the HMF one has 
$\langle p^2 \rangle = 2\varepsilon-1$, as can be easily derived from Eq. (\ref{ham0}). 
Then, the stability indicator $I$ as a function of the energy for the $q$-exponential 
family reads

\begin{equation}
I = 1-\frac{3-q}{2(5-3q)(2\varepsilon-1) } \;.
\end{equation}
Therefore, stability occurs for energies above $\varepsilon_{\rm q}$ 
\begin{equation} \label{threshold}
\varepsilon_{\rm q} = \frac{3}{4} +\frac{q-1}{2(5-3q)} \; .
\end{equation}
The stability diagram is exhibited in Fig.~2.
It is easy to verify that one recovers the known stability thresholds for the 
uniform and Gaussian distributions. We remark that Eq.~(\ref{threshold}) states that only 
finite-support distributions,  
corresponding to $q<1$, are stable below $\varepsilon_{\rm cr}$.
This agrees with numerical studies in the meta-equilibrium regimes of the HMF. 

\begin{figure}[ht]
\begin{center}
\includegraphics*[bb=90 480 550 720, width=0.7\textwidth]{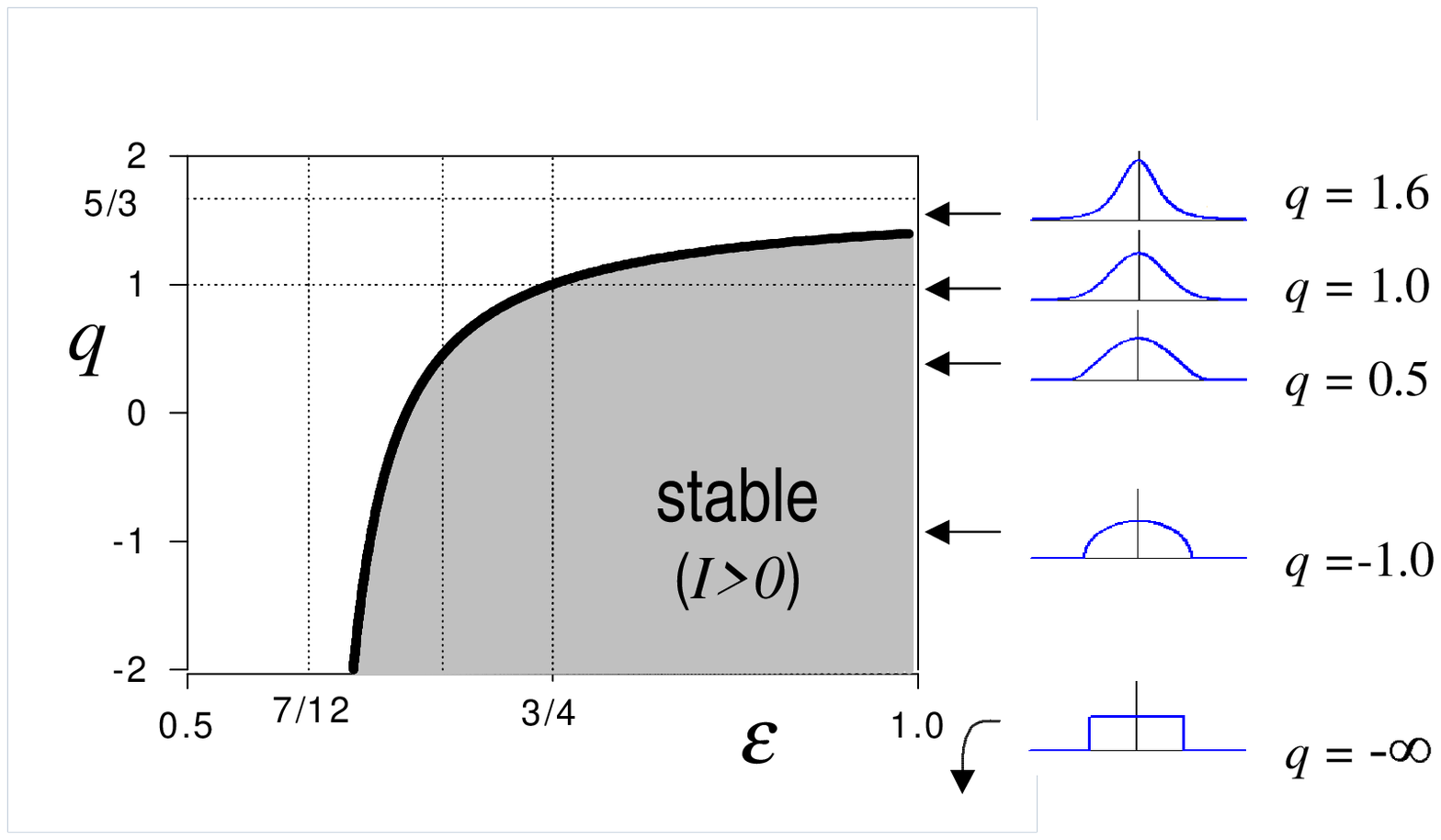}
\caption{\protect 
Stability diagram of the $q$-Gaussian ansatz for the momentum PDFs.} 
\end{center}
\label{fig:stability}
\end{figure}

We have also shown recently\cite{vlasov} that a similar analysis can be 
performed for a very simple family of functions exhibiting the basic structure of the observed 
$f(p)$, basically, a uniform distribution plus cosine. 
Fitting of numerical distributions leads to points in parameter space that fall close to the 
boundary of Vlasov stability, and exit the stability region for energies 
below the limiting value $\varepsilon \simeq 0.68$.
This result is confirmed when the stability criterion is applied to the 
discrete distributions arising from numerical simulations\cite{vlasov}, 
although for the discrete dynamics the magnetization is not strictly zero.

The stability index $I$ is positive for energies above $\varepsilon\simeq 0.68$. 
The fact that the stability indicator becomes negative below $\varepsilon\simeq 0.68$ 
signals the disappearance of the homogeneous metastable phase at that energy. 
{In fact, extrapolation of numerical simulations to the thermodynamic limit confirm this 
result. }
The present stability test only applies to homogeneous states. 
Strictly speaking, $m=0$ does not imply that the states 
are inhomogeneous. However, the sudden relaxation that leads to the 
present MESs mixes particles\cite{plr} in such a way 
that $m=0$ and spatial homogeneity are expected to be synonymous.  
Below $\varepsilon=0.68$, the measured distributions 
are evidently inhomogeneous ($m \neq 0$). In these cases, negative
stability refers to hypothetical homogeneous states having the measured $f(p)$. 

\section{Final remarks}

We have seen that, although our approach is valid in the continuum limit, 
it gives useful hints on the finite size dynamics. 
Of course, it can not predict complex details of the discrete dynamics. 
However, the present approach gives information on the violent initial 
relaxation from fully magnetized states, for sufficiently large system. 
It also explains the disappearance of homogeneous MESs below a certain 
energy observed by extrapolation of numerical simulations to the thermodynamic limit. 
Moreover, the identification of MESs with Vlasov solutions is also 
consistent with the fact that when the thermodynamic limit is taken before the 
limit $t\to\infty$, the system never relaxes to true equilibrium, remaining 
forever in a disordered state.

%%%%%%%%%%%%%%%%%%%%%%%%%%%
\section*{Acknowledgements}
%%%%%%%%%%%%%%%%%%%%%%%%%%%
C.A. is very grateful to the organizers  for the opportunity of participating 
of the nice meeting at 
the Ettore Majorana Foundation and Centre for Scientific Culture in Erice.

\end{document}